\documentclass[usenatbib,useAMS]{mn2e}
\usepackage{natbib,times}
\usepackage{longtable}
\usepackage{amssymb}
\usepackage{rotating}
\usepackage{epsfig}
\usepackage{amsmath}
\usepackage{graphicx}
\usepackage{makecell}

\def\kms{km~s$^{-1}$}

\begin{document}

\title[2MTF II. Parkes data]
{2MTF II. New Parkes 21-cm observations of 303 southern galaxies}

\author[T. Hong et al.]
{Tao~Hong~$^{1,2,3}$\thanks{E-mail: bartonhongtao@gmail.com}, 
Lister~Staveley-Smith~$^{2,3}$, 
Karen~L.~Masters~$^{4,5,6}$, 
\newauthor
Christopher~M.~Springob~$^{2,3,7}$, 
Lucas~M.~Macri~$^{8}$, 
B\"arbel~S.~Koribalski~$^{9}$, 
D.~Heath~Jones~$^{10}$,
\newauthor
Tom~H.~Jarrett~$^{11}$~and Aidan~C.~Crook~$^{12}$
\\
$^{1}$National Astronomical Observatories, Chinese Academy
  of Sciences, 20A Datun Road, Chaoyang District, Beijing 100012,
  China.\\
$^{2}$International Centre for Radio Astronomy Research,
  M468, University of Western Australia, Crawley, 35 Stirling Highway, WA 6009, Australia\\
$^{3}$ARC Centre of Excellence for All-sky Astrophysics
  (CAASTRO)\\
$^{4}$Institute for Cosmology and Gravitation, University 
of Portsmouth, Dennis Sciama Building, Burnaby Road, Portsmouth 
PO1 3FX\\
$^{5}$South East Physics Network (www.sepnet.ac.uk)\\
$^{6}$Harvard-Smithsonian Center for Astrophysics, 60 Garden Street, Cambridge, MA 02138, USA\\
$^{7}$Australian Astronomical Observatory, PO Box 915, North Ryde, NSW 1670 Australia\\
$^{8}$George P. and Cynthia Woods Mitchell Institute for Fundamental Physics and 
Astronomy, Department of Physics and Astronomy, Texas A\&M University, \\4242 TAMU, 
College Station, TX 77843, USA\\ 
$^{9}$CSIRO Astronomy \& Space Science, Australia Telescope National 
Facility, PO Box 76, Epping, NSW 1710, Australia\\ 
$^{10}$School of Physics, Monash University, Clayton, VIC 3800, Australia\\ 
$^{11}$Astronomy Department, University of Cape Town, Private Bag X3. 
Rondebosch 7701, Republic of South Africa\\
$^{12}$Microsoft Corporation, 1 Microsoft Way, Redmond, WA 98052\\
}

\date{Accepted  ... Received  ...}

\pagerange{\pageref{firstpage}--\pageref{lastpage}} \pubyear{}

\maketitle
\label{firstpage}
\begin{abstract}

We present new 21-cm neutral hydrogen (H\,{\sc i}) observations of spiral 
galaxies for the 2MASS Tully Fisher (2MTF) survey. Using the 64-m Parkes 
radio telescope multibeam system we obtain 152 high signal-to-noise H\,{\sc i} 
spectra from which we extract 148 high-accuracy ($< 5\%$ error) velocity widths
and derive reliable rotation velocities. The observed sample consists of 303 
southern ($\delta < -40\degr$) galaxies selected from the 2MASS Redshift Survey
(2MRS) with $K_s <11.25$ mag, $cz < 10,000$ km\,s$^{-1}$ and axis ratio $b/a 
< 0.5$. The H\,{\sc i} observations reported in this paper will be combined 
with new H\,{\sc i} spectra from the Green Bank and Arecibo telescopes, together
producing the most uniform 
Tully-Fisher survey 
ever constructed (in terms of sky coverage). In particular, due to its near infrared selection, 
2MTF will be significantly more complete at low Galactic latitude ($|b|<15^\circ$) 
and will provide a more reliable map of peculiar velocities in the local universe.
\end{abstract}

\section{Introduction}
In the local Universe, the galaxy distribution reveals large
structures such as walls, filaments and voids on scales up to 100 Mpc
\citep{lgh1986,jrs+2009,sdb+2012}. The gravitational effects exerted on
individual galaxies by this inhomogeneous distribution results in
peculiar (non-Hubble) motions that can be used to probe the
underlying mass distribution and constrain the cosmological models
\citep{elh+2006}. Much of our understanding of the local Universe
comes from optical sky surveys. However, infrared and 21-cm surveys are
increasingly important because of lower dust extinction and their
closer correspondence to stellar luminosity and total mass,
respectively.

An important application obtained from the combination of galaxy
photometry and H\,{\sc i} spectra is the infrared Tully-Fisher relation, which
is an empirical relation between the luminosity and rotational
velocity of spiral galaxies \citep{tf1977}. The near-infrared Tully-Fisher 
relation has increased precision over optical formulations \citep{ahm+1982} and 
can be calibrated via primary distance indicators such as Cepheids or the 
Tip of the Red Giant Branch \citep{tc2012}, it can be used to measure
redshift-independent distances of local spiral galaxies.  With these
redshift-independent distances, we can calculate the peculiar velocity
field.

In the last few decades, a number of Tully-Fisher surveys have been
conducted, including those described in
\citet{ghh+1997,smh+2007,tsk+2008}. These are typically limited by
source selection criteria and sky coverage. For instance, the SFI++
survey \citep[and references therein]{hgc+1999,hgs+1999,mshg2006,smh+2007}, 
which is the largest Tully-Fisher survey to
date, was selected optically in $I$-band and 
can only cover Galactic
latitudes $|b|>15^{\circ}$. The part of the sky not covered by
SFI++ is known as the Zone of Avoidance (ZoA) and is difficult to
observe because of the effects of dust and stellar crowding in the plane of our Galaxy.

The 2MASS Tully-Fisher Survey \citep[2MTF,][]{m2008, msh2008, hsm+2013} gets
around this by using infrared and 21-cm radio observations to improve
our knowledge and model of the mass distribution of the local
Universe. 2MTF is based on a source list selected from the 2 Micron
All-Sky Survey Extended Source Catalog \citep[2MASS XSC,][]{jcc+2000}, 
and combines high-quality infrared photometry
and 21-cm rotation widths for all bright inclined spirals in the 2MASS
Redshift Survey \citep[2MRS,][]{hmm+2012}. The final 2MTF sample is expected to 
contain about 3,000 high-quality H\,{\sc i} widths, including new observed H\,{\sc i} widths by our 
group using the Green Bank Telescope (GBT) and Parkes radio telescope, H\,{\sc i} widths from the ALFALFA survey 
\citep{ghk+2005, hgm+2011} and high quality archival H\,{\sc i} widths. 

In this paper, we present H\,{\sc i} observations of 303 southern 2MTF
galaxies using the Parkes radio telescope. We describe our observations and
data reduction processes in Section~\ref{sec:obs}.
In Section~\ref{sec:stat}, we discuss the statistical properties of
the data. Some notable detections are presented in
Section~\ref{sec:note}.  We give the summary in the last section.

\section{Observations and Data reduction}
\label{sec:obs}
The 2MTF survey aims to measure distances for all bright inclined
spirals in 2MRS. We selected galaxies from the 2MRS
catalog that met the following criteria: total $K_{s}$ magnitudes 
$K_{s} < 11.25$~mag, $cz<10,000$~\kms, and
axis ratio $b/a < 0.5$. In addition, we added some galaxies with 
$K_s < 11.75$~mag in order to increase the number of H\,{\sc i} detections 
at declinations south of $-40^{\circ}$. 
The target list contains $\sim 6,000$ 2MRS galaxies that 
meet our selection 
criteria. By 2006, when we made our observation 
plan, 40\% of the target galaxies already had archival
rotation width measurements for Tully-Fisher distances 
\citep[mainly from ][]{tbc+1998,shgk2005,tch+2005}, but with very
uneven sky coverage, especially in the southern hemisphere. 
To supplement these archival measurements, we
observed $\sim 1,000$ galaxies with $\delta>
-40^{\circ}$ with the Green Bank Telescope to peak flux limits
$S_{p} \geq 10$~mJy (Masters et al., in prep). For 
$\delta < -40^{\circ}$ only about 25\% of the 1018 selected 2MRS galaxies
had high-quality H\,{\sc i} width measurements already available. Of the
remaining 754 galaxies, 303 were deemed not to be confused in the 15
arcmin beam of the Parkes telescope, and were observed.

The southern galaxies were observed in six semesters between 2006 and
2012 (see Table~\ref{tab:obs} for more details) using the 21-cm
multibeam receiver \citep{swb+1996}. The multibeam correlator was used
with a bandwidth of 8 MHz, divided into 1024 channels, providing a 
channel spacing of $\sim1.6$~\kms. During the observation of each
galaxy, the band was centered on the 2MRS redshift of the target
galaxy. The observations were done in beam switching mode (MX mode)
using the 7 high-efficiency central beams of the receiver each with
two orthogonal linear polarizations.  In MX mode, the target galaxy
was tracked in turn with each beam. When a beam was not pointing at
the galaxy (off position), the data collected by this beam was used as a
reference spectrum for calibration of the on-galaxy spectrum.
\begin{table}
\caption[]{Details of Parkes observations}
\label{tab:obs}
\centering
\begin{tabular}{ccc}
\hline
\hline
Observing dates & Observing hours & Number of galaxies\\
\hline
2006 Nov 3 - Nov 12 & 80 & 68 \\
2007 May 20 - Jun 3 & 160 & 84 \\
2007 Nov 1 - Nov 7 & 55 & 22 \\
2007 Dec 5 - Dec 14 & 65 & 32 \\
2008 May 12 - May 22 & 146 & 115 \\
2008 Sep 24 - Oct 1 & 72 & 34 \\
2011 Oct 1 - Oct 6 & 40 &  33 \\
2012 Mar 11 - Mar 16 & 40 &  28 \\
\hline
\end{tabular}
\end{table}

Each galaxy was observed for a minimum of 35-min (i.e. each
of the 7 beams was on-source for 5-min), with the correlator
writing a spectrum every 5 seconds. After a preliminary data reduction, 
unless the observer estimated the signal-to-noise (S/N) of the galaxy 
H\,{\sc i} spectrum to be $\ga$10, the process was repeated. We define S/N
as the ratio of the peak H\,{\sc i} flux per channel divided by the rms
noise. 
Galaxies with profiles which were deemed too weak to reach that
S/N ratio in a reasonable time were not observed further.

The data were bandpass and Doppler corrected using \textsc{livedata}
\citep{bsbo+2001} with a MEDIAN estimator, all spectra were
corrected to the solar system barycenter. Gridding was done by
\textsc{gridzilla}, using a MEDIAN gridding algorithm. In
order to obtain identical H\,{\sc i} parameter measurements to the GBT 
observations (Masters et al., in prep.), we adopted the
same GBTIDL routines. Using 3-channel Hanning smoothing we obtained 
a velocity resolution of 3.3~\kms and rms of 2 - 17 mJy.

The main source of radio frequency interference (RFI) was the L3
beacon of the Global Position System (GPS) satellite near 1381 MHz
(equivalent to $cz \sim 8306$~\kms), which occurred approximately every 30-min. 
In order to avoid contaminating the H\,{\sc i} spectra of galaxies with 
velocities near this RFI signal, we reduced their on-source integration 
time from 35 to 21-min.

\begin{figure*}
\centering
\includegraphics[width=1.5\columnwidth, angle=-90]{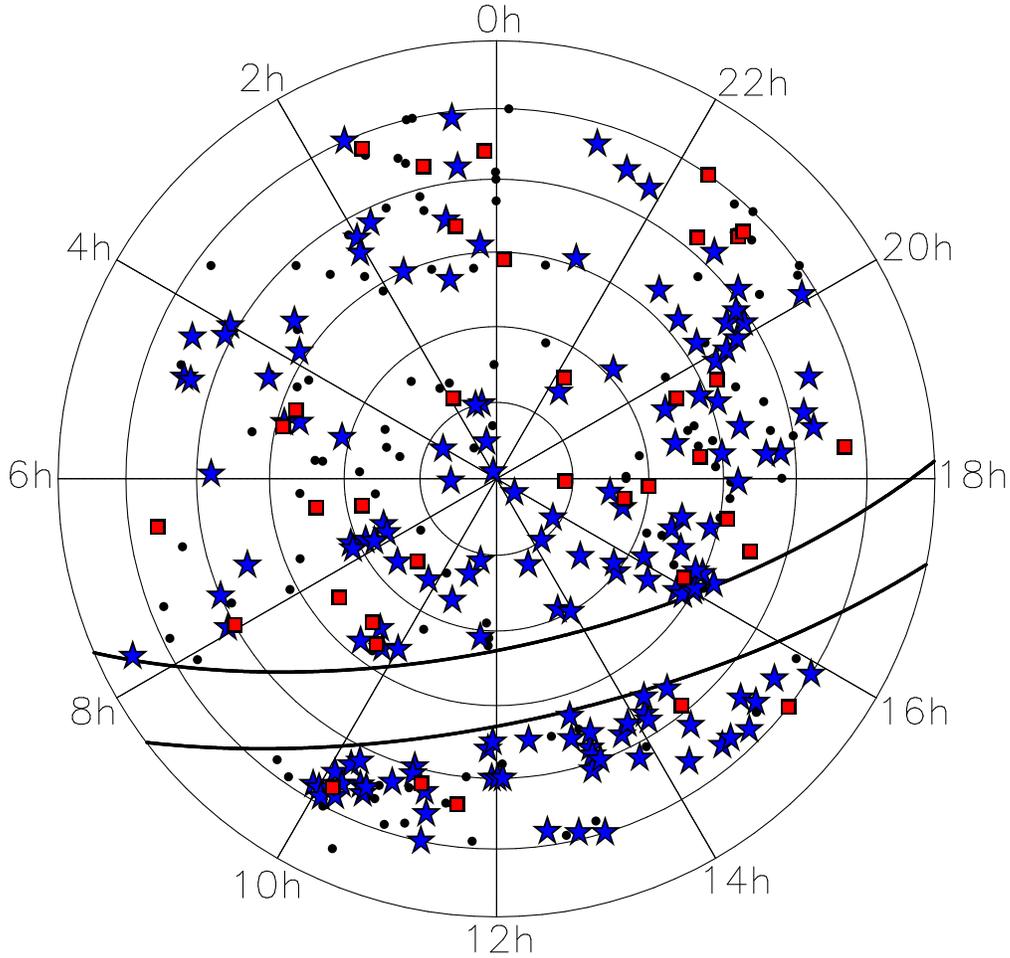}
\caption{The distribution of Parkes observed galaxies. The blue stars
  indicate the 152 well-detected galaxies; poorly-detected galaxies
  are plotted with red squares; the black dots are the non-detected
  galaxies. The thick lines trace the galactic latitudes $b=5^\circ$ 
  and $b=-5^\circ$. The center of the projection is at the south pole, 
  the latitude lines are plot in steps of $10^\circ$.}
\label{fig:sky_dis}
\end{figure*}

Of the 303 observed galaxies: 152 have spectra whose S/N and spectrum
profile are good enough for H\,{\sc i} parameter measurements; 36 were poorly
detected; and the remaining 115 galaxies were not detected. We report
the raw and corrected H\,{\sc i} parameters for the 152 well-detected galaxies
in this paper. Figure~\ref{fig:sky_dis} shows the sky distribution of all
Parkes observed galaxies.

\begin{figure*}
\centering
\includegraphics[width=0.95\textwidth]{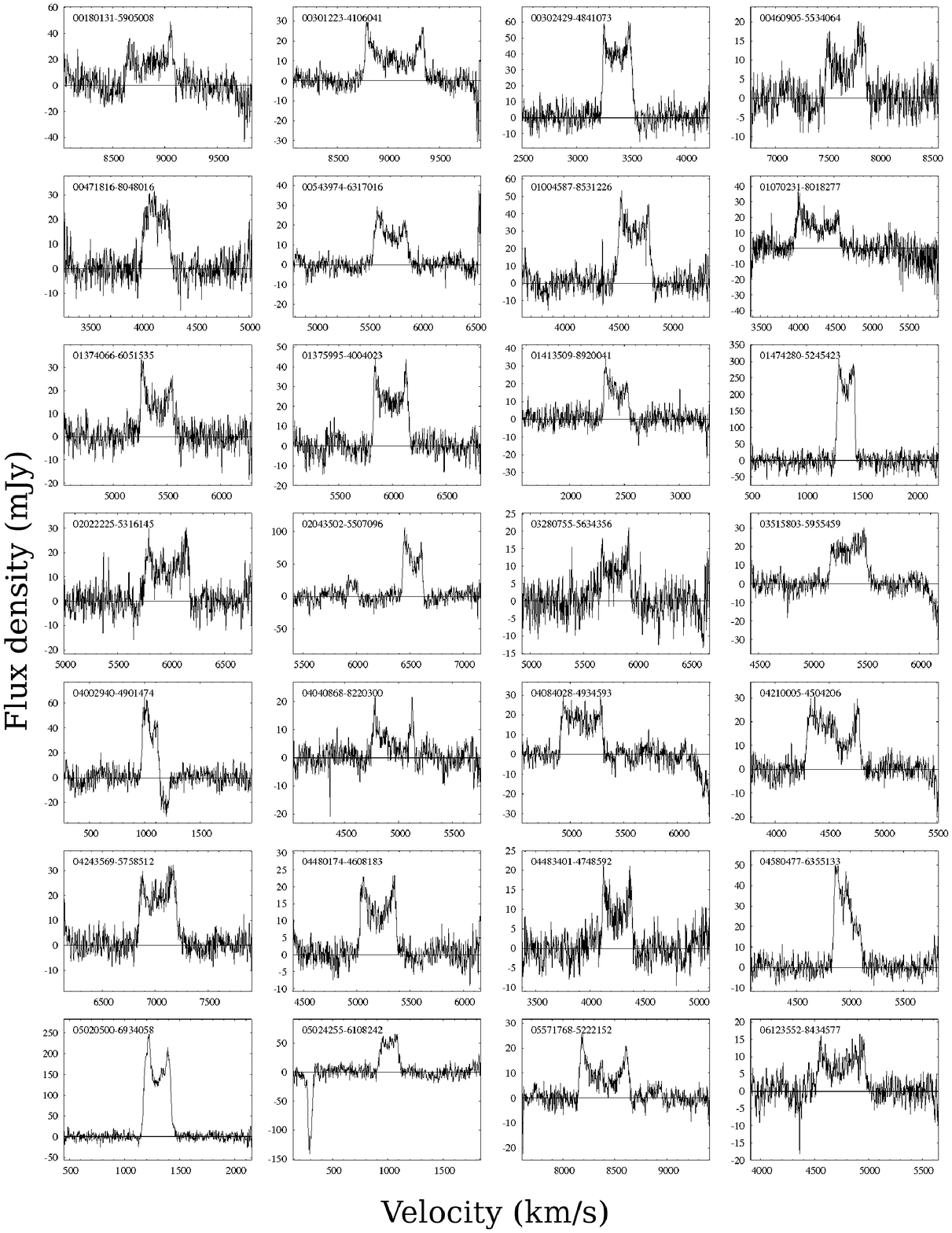}
\caption{28 H\,{\sc i} spectra of Parkes observed galaxies. The 2MASS name is
  given at the top of each spectrum. All spectra are baseline
  subtracted. All 152 spectra will be available in digital form at http://ict.icrar.org/2MTF.}
\label{fig:spectra}
\end{figure*}
\subsection{H\,{\sc i} parameter measurements}
\label{sec:data}
\subsubsection{Integrated line flux and errors}
We measured integrated line flux ($F_{obs}$) from the smoothed and
baseline-subtracted profiles. We manually marked the part of the spectrum
where the H\,{\sc i} emission line was present, and measured the integrated
line flux within these boundaries. A line-free region was also marked,
and the baseline (and noise $\sigma_{rms}$ in the spectrum) was
measured in this part of the spectrum.

We adopted a jackknife method to estimate the error on the H\,{\sc i} flux. For
each galaxy, we built 100 jackknife spectra by leaving out one percent
of the original data each time.  All 100 jackknife spectra were then measured
automatically using IDL routines, and the errors in H\,{\sc i} flux were taken
as:
\begin{equation}
\sigma_J=\left[ \frac{N-1}{N} \sum^{N}_{i=1} \left(f_{i}^{J}-\overline{f}^J \right)^2 \right]^{1/2},
\label{eq:sigma}
\end{equation}
where $N$ is the number of jackknife samples, $f_{i}^{J}$ is the
measurement for the $i$th jackknife spectrum, and $\overline{f}^J =
\frac{1}{N} \sum^{N}_{i=1} f_{i}^{J}$.  We give a detailed description
of jackknife error estimation in Appendix~\ref{sec:jack}.

%
%


\subsubsection{Systemic velocities and velocity widths}
\label{sec:velocity}
Systemic velocities and velocity widths are measured by 
selecting two points on opposite sides of the H\,{\sc i} emission profile.  The
velocity width is the velocity difference between the highest and
lowest velocities ($v_h$ and $v_l$, respectively), $W=v_h-v_l$.  The
systemic velocity is the average of the two velocities,
$V=(v_h+v_l)/2$. The choice of measurement algorithm can affect
accuracy, especially for the low S/N spectra. \citet{ksk+2004} used 
H\,{\sc i} widths measured at both 50\% and 20\% level of the peak flux
density ($W_{P50}$ and $W_{P20}$).  \citet{hgm+2011} adopted
algorithms which measured the widths at the 50\% level of each of the
two peaks ($W_{2P50}$). 

Separate from the question of which flux level at which to mark the two sides of the profile, there is the question of what method one uses to decide which {\it channel} corresponds to the given flux.  The most commonly used method involves either starting from the two peaks of the profile and moving outwards from the centre until one finds the first channel below the desired flux threshold, or starting from the outside of the line profile, and moving inwards until one finds a channel that exceeds the flux threshold.  This approach is adequate for most Tully-Fisher applications, because the S/N for most Tully-Fisher galaxies is sufficiently large that any noise along the sides of the line profile does not greatly complicate the width measurement.  Nevertheless, to guard against the possibility of noisy spectra perturbing our measurements of $v_h$ and $v_l$, we favour a width measurement algorithm that involves fitting straight lines to either side of the spectral line profile.

\citet{ghh+1997} presented a method (first implemented in the Arecibo Observatory ANALYZ-GALPAC data reduction software) which fits a straight line to either side of the emission profile between 15\% and 
85\% of the peak value ($f_p-\sigma_{rms}$), then selected the left
and right points at the 50\% level of the peak value from the fitted
lines ($W_{F50}$).  The method was later used by Springob et al. (2005), who updated the instrumental and velocity resolution correction.  This represents our ``favoured approach'' to measuring the line width.

We measured systemic velocities and velocity widths using a modified
version of the GBTIDL routine \textit{awv.pro} (see Masters et al. in
prep. who also use this).  This routine provides H\,{\sc i} parameter measurements using five
different algorithms: $W_{F50}$ is the width measured at 50\% of
the value of $f_p-\sigma_{rms}$ on a linear fit of both sides of the
profile; $W_{M50}$ is the width measured at 50\% of the mean flux
of the profile; $W_{2P50}$ is the width measured at 50\% of each
of the two $f_p-\sigma_{rms}$ values; $W_{P50}$ is the width measured
at 50\% of the $f_p-\sigma_{rms}$ value; and $W_{P20}$ is measured
at 20\% of the $f_p-\sigma_{rms}$ value.  $W_{F50}$ is the only one of these width measurements for which $v_h$ and $v_l$ are measured by fitting a line to either side of the profile.  The measurements of $v_h$ and $v_l$ for $W_{2P50}$ are made by starting at the peaks of the profile and moving outwards, while the corresponding measurements of $v_h$ and $v_l$ for $W_{M50}$, $W_{P50}$, and $W_{P20}$ are made by starting from the outside of the spectral line profile and moving inwards.

We report all five of these
widths here, so comparison with results in other databases can be made.
However, in this paper, we base our values for the final corrected
value ($W_c$) on the $W_{F50}$ measurement. As \citet{shgk2005} pointed
out, this reduces the dependence on the S/N of the spectra.

We applied four
corrections to $W_{F50}$ to obtain $W_c$: the instrumental correction,
the cosmological redshift correction, the correction for the turbulent
motions of H\,{\sc i} gas, and the correction for the inclination of the disk:
\begin{equation}
W_{c}=\left(\dfrac{W_{F50}-2\Delta_v\lambda}{1+z}-\Delta_t\right)\dfrac{1}{\sin{i}},
\label{width_corr}
\end{equation}
where $z$ is the redshift of the galaxy. $\Delta_v=3.3$~\kms\ is the velocity resolution of 
the spectrum. As given by \citet{shgk2005}, $\lambda$ is an empirically derived parameter 
for the instrumental correction 
that depends on the S/N and smoothing method (see \S 3.2.2
and Table 2 of \citet{shgk2005} for more information about this
correction).
$\Delta_t=6.5$~\kms\ is the correction for turbulent motions \citep{shgk2005}. Finally,
the inclination $i$ was estimated using the co-added axis ratio
($b/a$) from the 2MASS isophotal photometry by:
\begin{equation}
\cos^2{i}=\dfrac{(b/a)^2-q^2_0}{1-q^2_0},
\label{inclination}
\end{equation}
where we adopt $q_0=0.2$ as the intrinsic axis ratio for an edge-on spiral and set 
$\sin{i}=1$ for objects with $b/a$ below this value.

To estimate errors in the velocity parameters, we used a Monte-Carlo
method following \citet{dski+2005}. Every galaxy spectrum was smoothed 
by a Savitzky-Golay smoothing filter.  Fifty mock spectra were created for 
each galaxy by adding Poisson noise to the smoothed spectrum template, 
with the rms of the noise being equal to the rms of the original spectrum. 
Then the error was taken as the standard scatters of the
measurements of the fifty mock spectra.  We further discuss this
method and compare it with other error estimating methods in the
Appendix.

The errors on the $W_{F50}$ width are also corrected using
Equation 7 in \citet{ghh+1997}, which contains the uncertainties on
observations and all four corrections adopted for correcting the
widths. We report the corrected width error ($\epsilon_{wc}$) following the
corrected widths in the final data catalog.

\subsection{Catalog presentation}
We present the measured parameters of 152 well-detected galaxies in
Table~\ref{tab:data}. The contents of Table~\ref{tab:data} are as
follows.

Column (1). --- The 2MASS XSC ID name.

Column (2) and (3). --- Right ascension (RA) and declination (DEC) in
the J2000.0 epoch from the 2MASS XSC.

Column (4). --- The heliocentric redshift $V_{2MRS}$ from the 2MRS (\kms).

Column (5). --- The morphological type code following the RC3 system. 
Classification comes from the 2MRS.

Column (6). --- Co-added axis ratio ($b/a$) from the 2MASS XSC.

Column (7). --- The observed integrated 21-cm H\,{\sc i} line flux $F_{obs}$ 
(Jy~\kms).


Column (8). --- The uncertainty $\epsilon_F$ of the observed integrated H\,{\sc i} line flux (Jy~\kms).

Column (9). --- The heliocentric systemic velocity $V_{HI}$ of the 
H\,{\sc i} emission profile, generated by the fitting algorithm discussed in 
\S~\ref{sec:velocity}, taken as the midpoint of the velocity at 50\% 
level of $f_p-rms$, in \kms.

Columns (10-14). --- The velocity widths of the H\,{\sc i} line in \kms, using 
the five measuring algorithms discussed in \S~\ref{sec:velocity}. The 
widths are $W_{F50}$, $W_{M50}$, $W_{2P50}$, $W_{P50}$ and $W_{P20}$
respectively.

Columns (15-29). --- The observing error of five widths, estimated by 
the Monte-Carlo method. $\epsilon_{F50}$, $\epsilon_{M50}$, 
$\epsilon_{2P50}$, $\epsilon_{P50}$ and $\epsilon_{P20}$ 
respectively, also in \kms.

Column (20). --- The corrected velocity width $W_c$, in \kms, which 
accounts for all four corrections discussed in \S~\ref{sec:velocity}. 
The correction is applied to $W_{F50}$ only.

Column (21). --- The uncertainty $\epsilon_{Wc}$ of the corrected 
velocity (\kms).

Column (22). --- Peak signal-to-noise ratio of the H\,{\sc i} line, $\textrm{S/N}=f_p/\sigma_{rms}$

Column (23). --- Velocity width instrumental correction parameter, $\lambda$.
\begin{table*}
\caption[]{~H\,{\sc i} Parameters of well-detected galaxies}
\small
\label{tab:data}
\resizebox{\linewidth}{!}{
\begin{tabular}{lrrrcrrrrrrrrrrrrrrrrrc}
\hline\hline
\multicolumn{1}{c}{2MASX ID} &\multicolumn{1}{c}{RA}  & \multicolumn{1}{c}{DEC} &\multicolumn{1}{c}{$V_{2MRS}$} &\multicolumn{1}{c}{$T$} &\multicolumn{1}{c}{$b/a$} &\multicolumn{1}{c}{$F_{obs}$} &\multicolumn{1}{c}{$\epsilon_F$} &\multicolumn{1}{c}{$V_{HI}$}& \multicolumn{1}{c}{$W_{F50}$} & \multicolumn{1}{c}{$W_{M50}$} & \multicolumn{1}{c}{$W_{2P50}$} & \multicolumn{1}{c}{$W_{P50}$} & \multicolumn{1}{c}{$W_{P20}$} & \multicolumn{1}{c}{$\epsilon_{F50}$} & \multicolumn{1}{c}{$\epsilon_{M50}$} & \multicolumn{1}{c}{$\epsilon_{2P50}$} & \multicolumn{1}{c}{$\epsilon_{P50}$} & \multicolumn{1}{c}{$\epsilon_{P20}$} & \multicolumn{1}{c}{$W_{c}$} & \multicolumn{1}{c}{$\epsilon_{Wc}$} & \multicolumn{1}{c}{S/N} & $\lambda$\\
\cline{9-21}
\multicolumn{1}{c}{-~-} &\multicolumn{2}{c}{[deg] (J2000)} &\multicolumn{1}{c}{[\kms]} &\multicolumn{1}{c}{-~-} &\multicolumn{1}{c}{-~-} &\multicolumn{2}{c}{[Jy~\kms]} &\multicolumn{13}{c}{[\kms]} & \multicolumn{1}{c}{-~-} & \multicolumn{1}{c}{-~-}  \\

\multicolumn{1}{c}{(1)} & \multicolumn{1}{c}{(2)} & \multicolumn{1}{c}{(3)} & \multicolumn{1}{c}{(4)} & \multicolumn{1}{c}{(5)} & \multicolumn{1}{c}{(6)} & \multicolumn{1}{c}{(7)} & \multicolumn{1}{c}{(8)} & \multicolumn{1}{c}{(9)} & \multicolumn{1}{c}{(10)} & \multicolumn{1}{c}{(11)} & \multicolumn{1}{c}{(12)} & \multicolumn{1}{c}{(13)} & \multicolumn{1}{c}{(14)} & \multicolumn{1}{c}{(15)} & \multicolumn{1}{c}{(16)} & \multicolumn{1}{c}{(17)} & \multicolumn{1}{c}{(18)} & \multicolumn{1}{c}{(19)} & \multicolumn{1}{c}{(20)} & \multicolumn{1}{c}{(21)}& \multicolumn{1}{c}{(22)} & \multicolumn{1}{c}{(23)}\\
\hline
00180131-5905008	&    4.5056	 & -59.0836	&8924	&4	&0.30	&8.92	&0.87	&8856	&452	&488	&444	&433	&488	&13	&13	&9	&25	&14	&442	&13.1	&6.45	&0.167\\
00301223-4106041	&    7.5509	 & -41.1012	&8988	&6	&0.14	&8.53	&1.31	&9066	&589	&601	&591	&582	&606	&3	&6	&4	&4	&7	&563	&3.8 	&10.12	&0.321\\
00302429-4841073	&    7.6013	 & -48.6853	&3352	&0	&0.28	&11.78	&0.40	&3378	&276	&283	&276	&271	&296	&2	&3	&4	&4	&5	&269	&3.6 	&11.26	&0.357\\
00460905-5534064	&   11.5376	 & -55.5685	&7660	&1	&0.38	&4.00	&0.56	&7676	&396	&403	&397	&386	&409	&10	&6	&7	&5	&7	&401	&11.4	&6.11	&0.149\\
00471816-8048016	&   11.8259	 & -80.8005	&4138	&1	&0.36	&6.03	&0.25	&4133	&260	&268	&252	&246	&281	&10	&9	&9	&6	&11	&260	&10.7	&8.43	&0.258\\
00543974-6317016	&   13.6656	 & -63.2838	&5662	&4	&0.40	&5.94	&0.35	&5710	&324	&335	&329	&320	&355	&4	&6	&5	&5	&9	&330	&6.5 	&10.78	&0.342\\
01004587-8531226	&   15.1912	 & -85.5229	&4657	&6	&0.16	&10.07	&0.60	&4657	&300	&320	&297	&291	&338	&3	&8	&5	&5	&10	&286	&3.8 	&10.82	&0.343\\
01070231-8018277	&   16.7593	 & -80.3077	&5047	&-2	&0.32	&10.66	&0.58	&4286	&597	&628	&593	&592	&629	&7	&10	&10	&12	&11	&598	&9.5 	&8.76	&0.271\\
01374066-6051535	&   24.4194	 & -60.8649	&5425	&5	&0.40	&5.26	&0.35	&5411	&310	&310	&308	&306	&330	&3	&5	&4	&4	&7	&316	&6.2 	&8.31	&0.254\\
01375995-4004023	&   24.4998	 & -40.0673	&5985	&6	&0.30	&8.41	&0.88	&5987	&327	&335	&329	&325	&344	&3	&3	&3	&3	&6	&320	&4.3 	&11.03	&0.350\\
01413509-8920041	&   25.4056	 & -89.3345	&2429	&4	&0.40	&3.89	&0.54	&2429	&229	&229	&226	&216	&232	&4	&4	&5	&5	&7	&233	&5.6 	&8.32	&0.254\\
02022225-5316145	&   30.5928	 & -53.2707	&5884	&3	&0.24	&6.44	&0.55	&5971	&411	&466	&417	&417	&466	&8	&14	&6	&7	&14	&399	&8.7 	&8.04	&0.242\\
02043502-5507096	&   31.1458	 & -55.1193	&6293	&1	&0.48	&12.54	&0.32	&6529	&188	&194	&186	&185	&199	&2	&3	&2	&2	&3	&195	&5.2 	&15.70	&0.395\\
03280755-5634356	&   52.0315	 & -56.5766	&5797	&0	&0.42	&2.72	&0.37	&5800	&274	&287	&280	&265	&288	&6	&7	&13	&6	&8	&282	&8.1 	&5.56	&0.116\\
03515803-5955459	&   57.9918	 & -59.9294	&5301	&1	&0.44	&7.25	&0.56	&5338	&349	&355	&346	&339	&390	&4	&13	&5	&6	&11	&364	&8.3 	&9.47	&0.298\\
04002940-4901474	&   60.1225	 & -49.0299	&1059	&1	&0.44	&6.34	&0.37	&1051	&145	&149	&147	&142	&162	&3	&3	&3	&3	&3	&147	&4.7 	&12.43	&0.391\\
04040868-8220300	&   61.0360	 & -82.3417	&4893	&1	&0.34	&2.49	&0.75	&4950	&381	&403	&383	&377	&403	&4	&7	&5	&5	&7	&382	&5.8 	&7.01	&0.195\\
04084028-4934593	&   62.1678	 & -49.5831	&5421	&98	&0.42	&7.14	&0.55	&5110	&388	&396	&392	&386	&399	&3	&5	&4	&4	&5	&402	&7.7 	&7.46	&0.216\\
04210005-4504206	&   65.2502	 & -45.0724	&4639	&3	&0.24	&9.00	&1.23	&4540	&493	&509	&490	&490	&518	&4	&5	&5	&4	&9	&481	&4.8 	&10.46	&0.332\\
04243569-5758512	&   66.1487	 & -57.9809	&7057	&4	&0.48	&8.29	&0.74	&7027	&359	&364	&361	&352	&399	&5	&11	&5	&5	&11	&381	&10.3	&9.17	&0.287\\
04480174-4608183	&   72.0071	 & -46.1385	&5276	&3	&0.18	&5.42	&0.46	&5203	&342	&346	&342	&342	&365	&3	&5	&4	&4	&8	&327	&3.5 	&9.61	&0.303\\
04483401-4748592	&   72.1417	 & -47.8164	&4316	&6	&0.48	&3.22	&0.38	&4255	&281	&288	&282	&276	&297	&4	&6	&4	&5	&7	&301	&8.4 	&7.37	&0.212\\
04580477-6355133	&   74.5200	 & -63.9204	&4946	&10	&0.40	&8.42	&0.28	&4980	&264	&261	&267	&173	&275	&4	&3	&4	&22	&4	&268	&6.4 	&14.72	&0.395\\
05020500-6934058	&   75.5209	 & -69.5683	&1300	&5	&0.30	&43.41	&0.44	&1302	&244	&259	&243	&241	&264	&1	&1	&1	&1	&2	&240	&2.8 	&33.63	&0.395\\
05024255-6108242	&   75.6773	 & -61.1401	&1014	&8	&0.48	&9.70	&0.26	&1014	&177	&186	&175	&175	&200	&4	&5	&6	&6	&6	&188	&6.5 	&8.86	&0.275\\
05571768-5222152	&   89.3237	 & -52.3709	&8505	&6	&0.18	&5.44	&1.33	&8400	&477	&495	&478	&469	&495	&3	&7	&5	&5	&7	&455	&3.9 	&10.61	&0.337\\
06123552-8434577	&   93.1474	 & -84.5827	&4784	&3	&0.26	&4.16	&0.62	&4756	&444	&461	&443	&442	&469	&5	&7	&19	&7	&6	&435	&5.7 	&6.84	&0.187\\
07174094-5558355	&  109.4206	 & -55.9765	&8772	&4	&0.40	&3.24	&0.56	&8770	&429	&435	&433	&420	&435	&9	&6	&10	&7	&6	&437	&11.3	&5.55	&0.116\\
07284622-7436114	&  112.1926	 & -74.6032	&6205	&4	&0.22	&2.68	&0.38	&6253	&409	&422	&406	&405	&422	&5	&5	&5	&7	&5	&394	&5.6 	&6.49	&0.169\\
07351844-5014596	&  113.8269	 & -50.2500	&1200	&5	&0.38	&15.03	&0.26	&1202	&90	&115	&89	&88	&125	&2	&3	&3	&3	&4	&85	&2.8 	&22.67	&0.395\\
07361230-6947467	&  114.0514	 & -69.7963	&1499	&5	&0.48	&2.16	&0.14	&1366	&108	&101	&108	&87	&114	&7	&6	&9	&9	&6	&111	&8.2 	&6.87	&0.189\\
07425487-7113095	&  115.7285	 & -71.2193	&8428	&5	&0.46	&3.32	&0.85	&8395	&446	&513	&427	&443	&511	&9	&11	&10	&12	&12	&469	&13.8	&9.15	&0.286\\
07441185-3548317	&  116.0495	 & -35.8088	&2879	&5	&0.24	&26.47	&0.39	&2889	&277	&300	&283	&262	&302	&1	&2	&3	&2	&2	&267	&2.7 	&28.80	&0.395\\
07445959-7431008	&  116.2485	 & -74.5169	&4832	&5	&0.40	&3.84	&0.33	&4836	&330	&339	&332	&327	&340	&6	&5	&4	&5	&6	&339	&7.8 	&6.26	&0.157\\
07472880-6942350	&  116.8700	 & -69.7097	&4043	&6	&0.28	&2.88	&0.44	&3983	&328	&332	&333	&310	&335	&10	&7	&10	&12	&7	&322	&10.5	&5.53	&0.114\\
07501902-7252285	&  117.5794	 & -72.8746	&5261	&3	&0.46	&9.34	&0.48	&5213	&367	&385	&369	&369	&397	&5	&9	&6	&6	&9	&389	&9.7 	&8.29	&0.252\\
07581500-4951050	&  119.5625	 & -49.8514	&1119	&6	&0.20	&100.73	&0.79	&1117	&314	&323	&313	&310	&330	&1	&1	&1	&1	&2	&304	&2.2 	&31.04	&0.395\\
08422422-7352468	&  130.6005	 & -73.8797	&5097	&5	&0.30	&5.73	&0.41	&5099	&321	&324	&323	&316	&325	&5	&4	&5	&4	&5	&317	&5.7 	&5.77	&0.129\\
09232812-6252562	&  140.8672	 & -62.8823	&4562	&20	&0.30	&8.43	&0.48	&4561	&278	&290	&274	&273	&304	&4	&6	&4	&5	&7	&273	&4.5 	&9.58	&0.302\\
09314981-6541445	&  142.9574	 & -65.6957	&4698	&4	&0.44	&8.98	&0.53	&4699	&373	&391	&369	&369	&394	&5	&5	&5	&5	&7	&391	&9.2 	&8.53	&0.262\\
\hline
\multicolumn{23}{l}{Table~\ref{tab:data} is available in its entirety online. A portion is shown here for guidance regarding its form and content.}
\end{tabular}}
\footnotesize

\end{table*}

\section{Data Characteristics}
\label{sec:stat}
The sky distribution for all 152 well-detected galaxies is shown in
Figure~\ref{fig:sky_dis} by blue stars.  As discussed by \citet{smh+2007}, the
SFI++ catalog leaves a large gap near the Galactic plane, with only a
few galaxies included in the region $|b|<15^\circ$. Thus, a
significant number of our Parkes observations were focused on this low
Galactic latitude area where we provide 69 high-accuracy H\,{\sc i}
measurements. Note that even in the NIR, dust obscuration and 
stellar crowding still leave a small
ZoA at Galactic latitudes $|b|<5^\circ$.
%
%

\begin{figure}
\centering
\includegraphics[width=0.63\columnwidth]{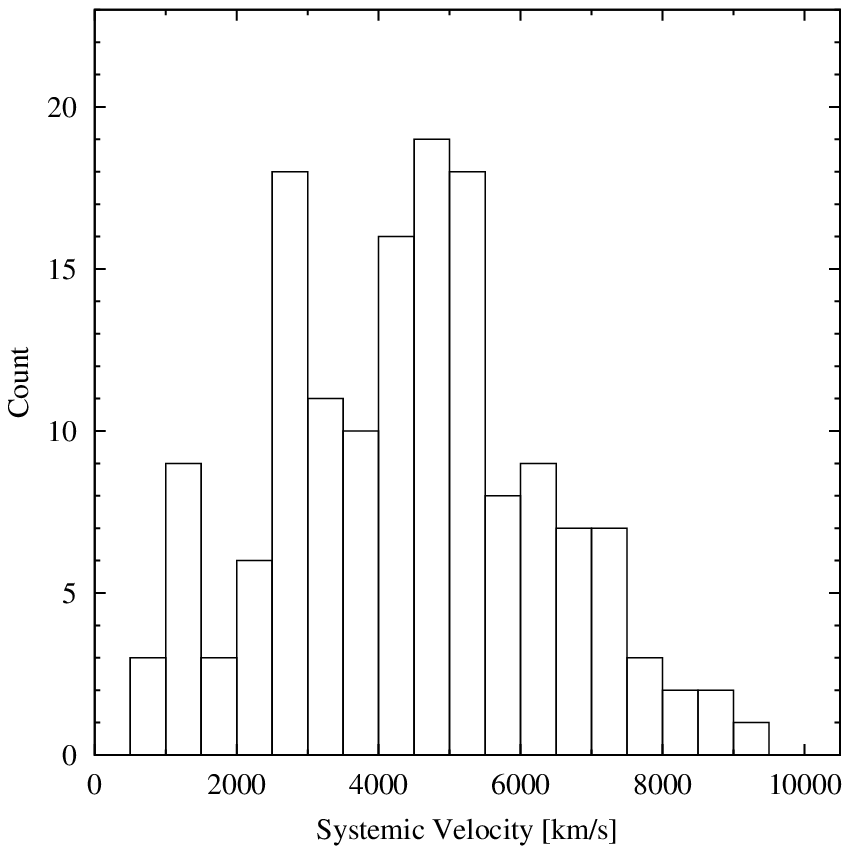}
\caption{The systemic velocity distributions of the 152 high quality
  Parkes galaxies, in bins of width $500$~\kms.}
\label{fig:his_vel}
\end{figure}

\begin{figure}
\centering
\includegraphics[width=0.60\columnwidth, angle=-90]{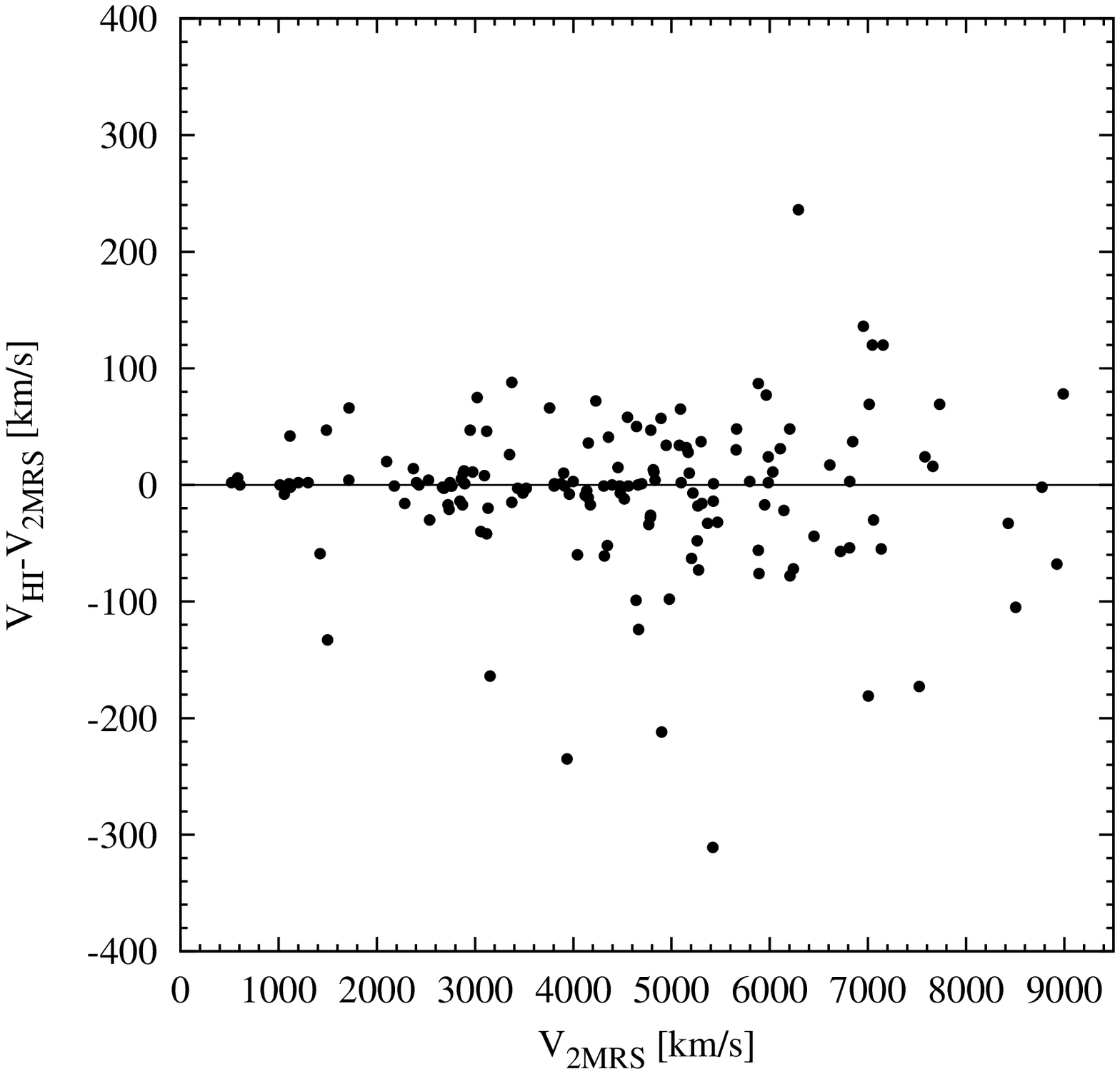}
\caption{$V_{HI}$-$V_{2MRS}$ versus $V_{2MRS}$ plot of 152 galaxies. The scatter about the line is 77~\kms.}
\label{fig:comp_velo}
\end{figure}

Figure~\ref{fig:his_vel} shows the H\,{\sc i} systemic velocity distribution
of the 152 galaxies.  As discussed in Section~\ref{sec:obs}, we
limited our sample to $cz < 10,000 $~\kms to get better H\,{\sc i}
profiles. Of the 152 measured systemic velocities, 121 (about 80\%) are less
than 6,000~\kms, with a mean value of 152 systemic velocities
$\overline{V_{HI}}=4433$~\kms. The highest-velocity galaxy is at $cz =
9066$~\kms, and the nearest one is at $cz = 524$~\kms.
The distribution of the differences between 2MRS and H\,{\sc i} systemic velocities is shown in
Figure~\ref{fig:comp_velo}.
%

%
%

\begin{figure}
\centering
\includegraphics[width=0.63\columnwidth]{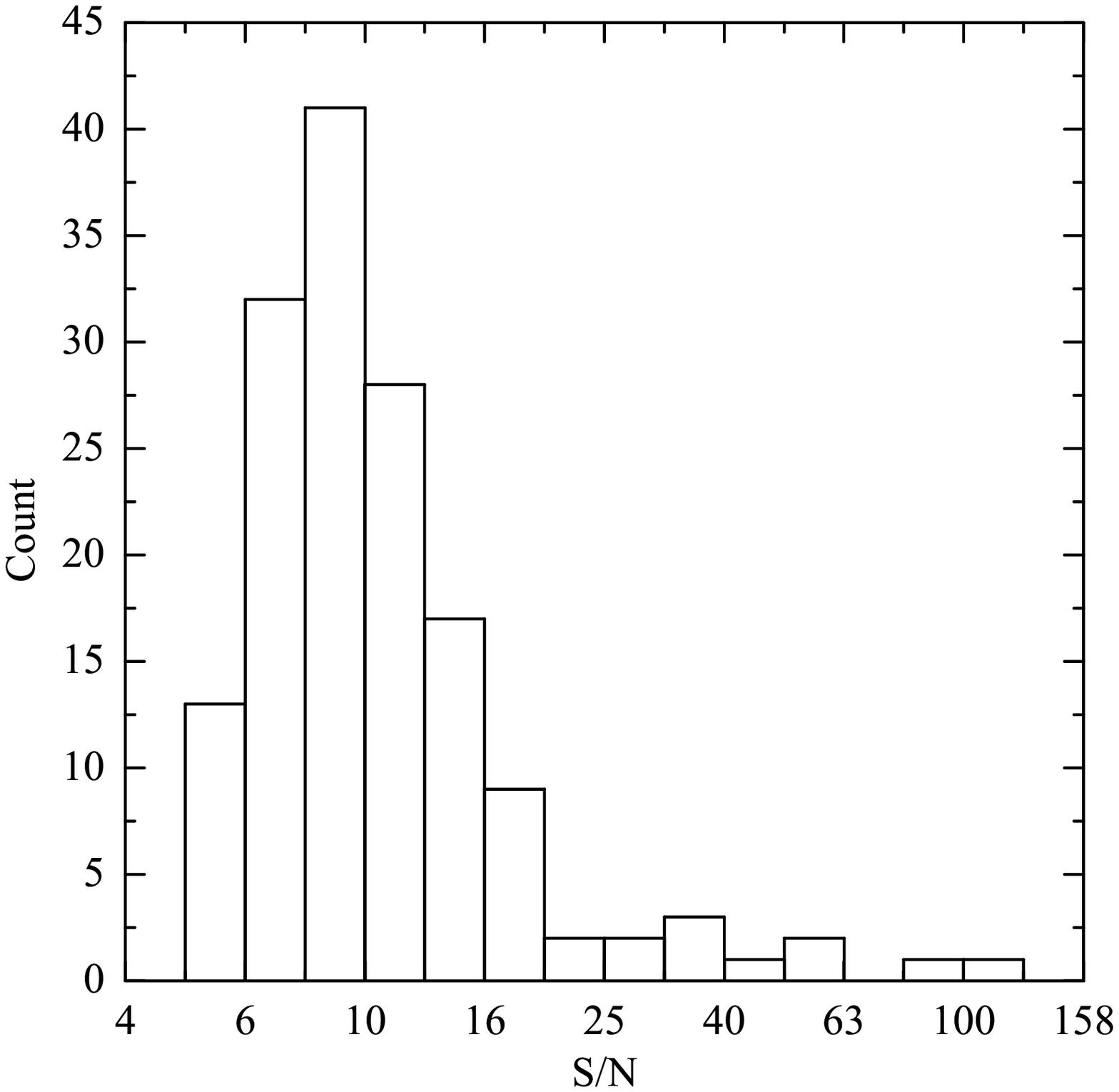}
\caption{The distribution of peak Signal-to-Noise ratio of the 152
  high quality Parkes galaxies, in bins of width 0.1 dex in logarithmic space.}
\label{fig:his_snr}
\end{figure}

\begin{figure}
\centering
\includegraphics[width=0.63\columnwidth]{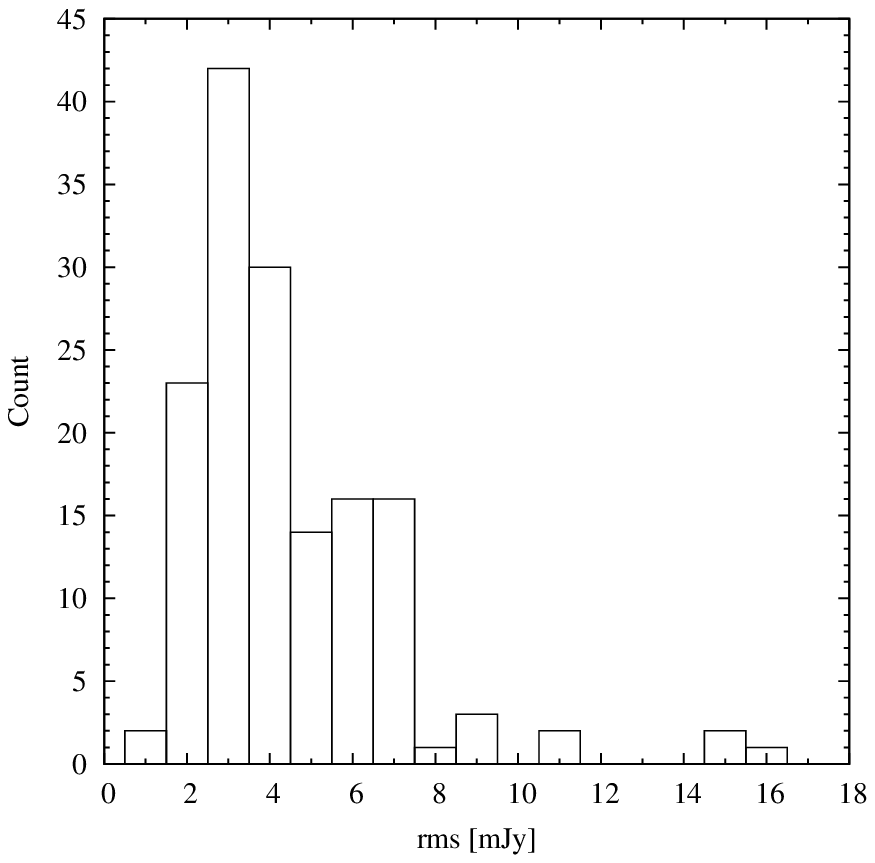}
\caption{The distribution of rms of the 152 high quality Parkes
  galaxies, in bins of width 1~mJy.}
\label{fig:his_rms}
\end{figure}

\begin{figure}
\centering
\includegraphics[width=0.63\columnwidth]{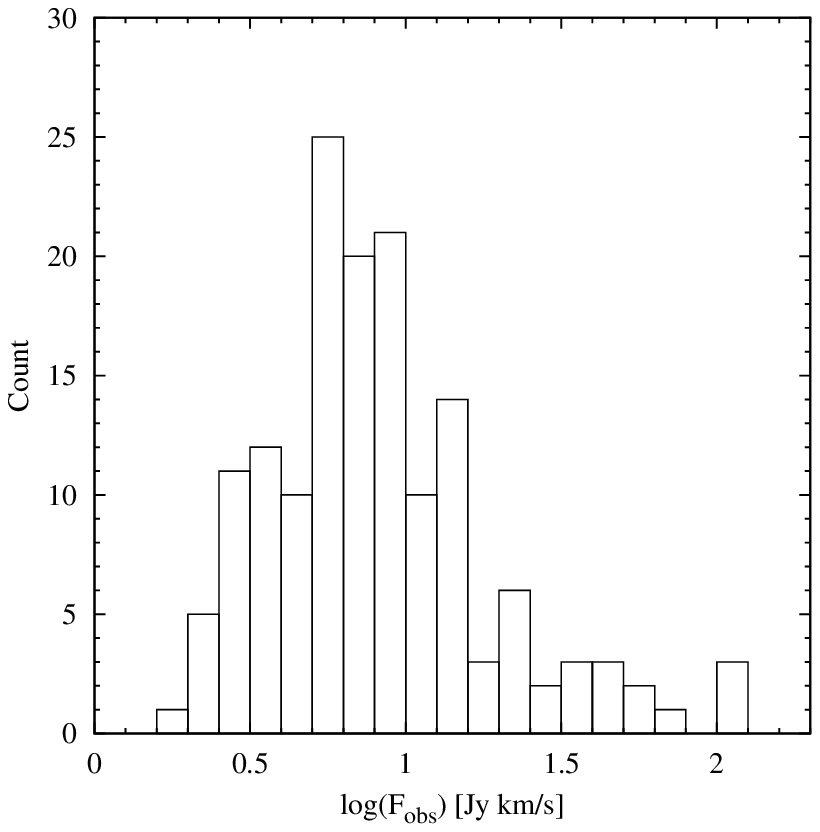}
\caption{The distribution of integrated H\,{\sc i} flux for 152 galaxies, in
  bins of width 0.1 dex.}
\label{fig:his_flux}
\end{figure}

The distributions of the peak signal-to-noise ratio
($\textrm{S/N}=f_p/\sigma_{rms}$) and rms are shown in Figure~\ref{fig:his_snr} and Figure~\ref{fig:his_rms} respectively.  All
galaxies have a $\textrm{S/N} > 5$, and 66 S/Ns are larger than
10. Generally, for the IDL routines we used to reduce the Parkes H\,{\sc i}
data, a S/N larger than five appears to be sufficient to measure an
accurate width.
Figure~\ref{fig:his_flux} shows the distribution of the observed
integrated H\,{\sc i} flux $F_{obs}$.  Compared to the catalog presented by
\citet{shgk2005}, our sample detects galaxies with larger integrated
H\,{\sc i} flux, because of the source selection criteria and the limit of
telescope sensitivity. As indicated by the Figure~\ref{fig:his_flux},
the distribution of H\,{\sc i} flux shows a peak at $\sim 7$~Jy~\kms, with a
mean value of $\overline{F_{obs}}=12.4$~Jy~\kms.

\begin{figure*}
\centering
\includegraphics[width=0.63\columnwidth]{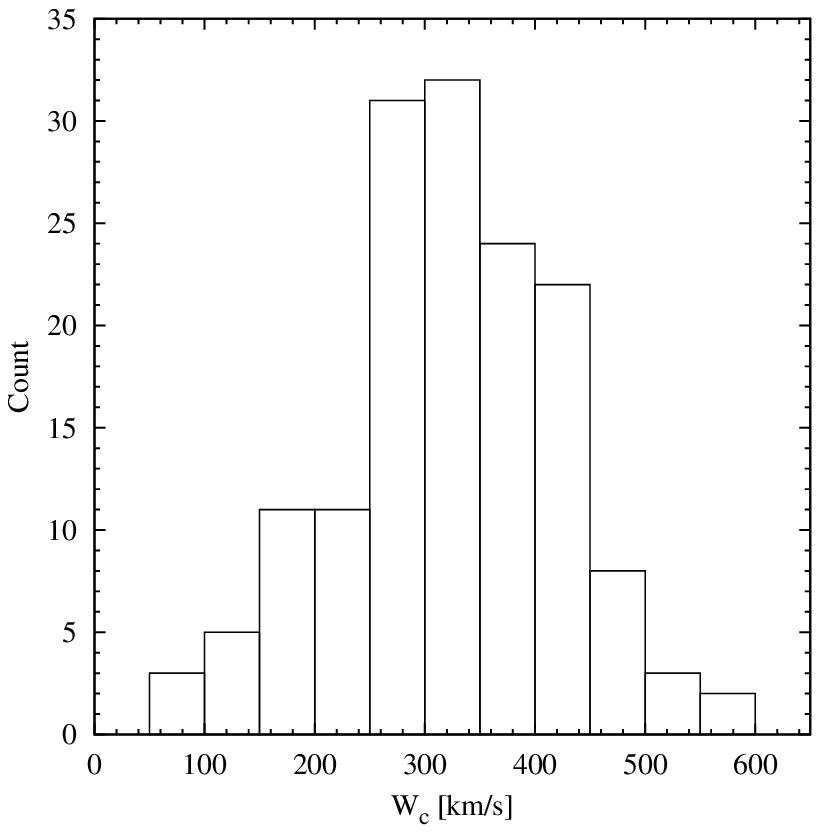}
\hspace{0.1cm}
\includegraphics[width=0.64\columnwidth]{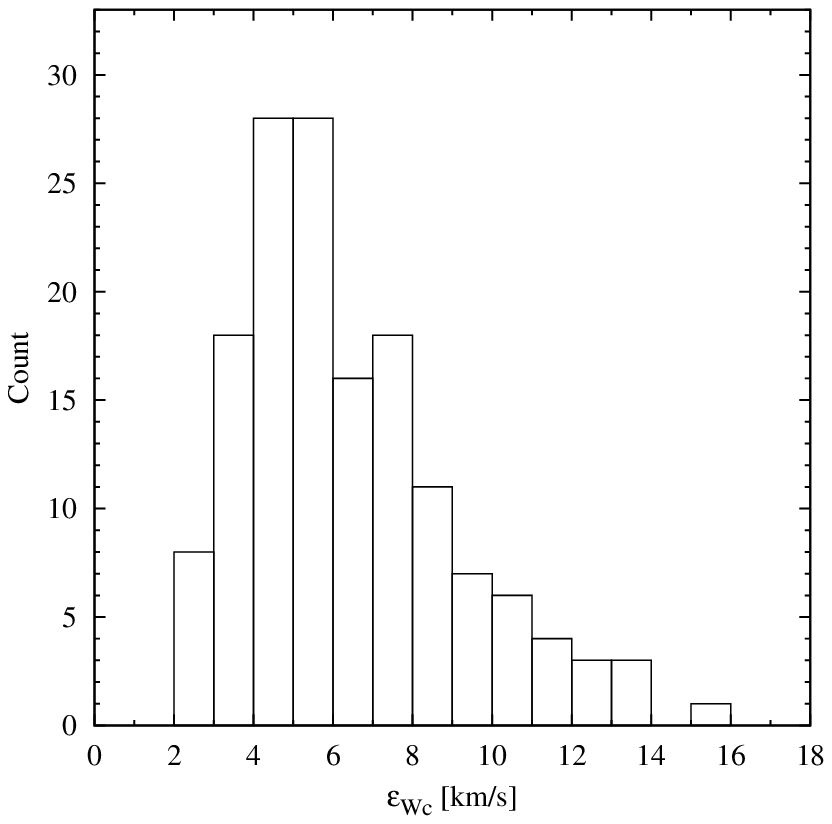}
\vspace{0.3cm}
\includegraphics[width=0.64\columnwidth]{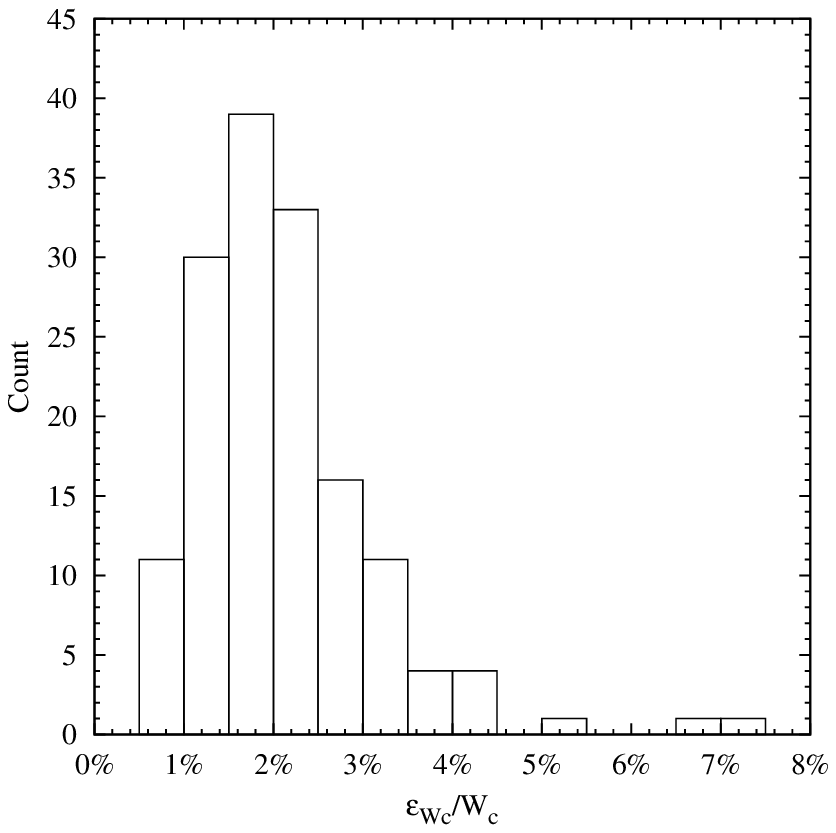}
\caption{The distributions of corrected widths, absolute errors of
  corrected widths and relative errors, in bins of
  50~\kms, 1~\kms and 0.5\% respectively.}
\label{fig:his_wc}
\end{figure*}
Finally, we show the histograms for the corrected widths $W_{c}$ and
the errors of corrected widths $\epsilon_{Wc}$ in
Figure~\ref{fig:his_wc}.  In comparison to the catalog of the 1000
brightest HIPASS galaxies, our catalog includes faster rotators, again
mainly because of the selection criteria. For the 2MTF project, whose
final goal is estimating the redshift-independent distances for spiral
galaxies, the accuracy of H\,{\sc i} width is one of the most important target
parameters. 
All but four of the galaxies have relative width errors below 5\%: 
the profiles of 2MASX 07361230-6947467 and 12541830-4149141 have 
$\textrm{S/N} \sim 6$; 2MASX 14242324-8027573 has a well-measured profile ($\textrm{S/N} > 10$) 
but its slow rotation ($W_{c}= 74$~\kms) amplifies the relative error; 
2MASX 01474280-5245423 has an excellent profile ($\textrm{S/N} \sim 18$) but 
its unfavorable inclination ($b/a = 0.8$) causes a very large uncertainty in the width.
As the latter galaxy has $b/a > 0.5$, it is eliminated from further cosmological analysis.
\\

%
%

\section{Notable Detections}
\label{sec:note}
\subsection{Discrepant Velocities}
We compared our derived H\,{\sc i} systemic velocities with those listed in 
the NASA/IPAC Extragalactic Database (NED) and found 5 objects that 
are discrepant by more than $ 3 \sigma$, as listed below:

(i) 2MASX 01070231-8018277: NED
prefers $cz = 5047 \pm 21$~\kms \citep{lv1989}, but also lists $cz = 4145 \pm 27$~\kms
\citep{wbw+2003} and $cz = 4249 \pm 27$~\kms \citep{cpd+1991}. We
determined $V_{HI} = 4286 \pm 4$~\kms, thus confirming
the alternate velocities.

(ii) 2MASX 18363723-4703153: NED prefers $cz = 7005 \pm 29$~\kms 
\citep{hmm+2012}.  We determined $V_{HI}=6824 \pm 1$~\kms, in 
agreement with $cz = 6857 \pm 45$~\kms from the 6dF Galaxy Survey \citep[6dFGS,][]{jrs+2009}.

(iii) 2MASX 20453927-5826591: NED
prefers $cz = 6954 \pm 45$~\kms \citep{jrs+2009}, but also lists 
$cz=7105 \pm 89$~\kms from the 2dF Galaxy Redshift Survey catalog (2dFGRS) . 
We determined 
$V_{HI}=7090 \pm 2$~\kms, in better agreements with the 2dFGRS value.

(iv) 2MASX 16375253-6448486: NED prefers $cz=4900 \pm 70$~\kms \citep{ncpp1997}. 
We determined $V_{HI} = 4688 \pm 1$~\kms which agrees with the 
HIPASS velocity \citep{ddr+2005} of $cz=4693$~\kms.

(v) 2MASX 02043502-5507096: NED prefers $cz = 6293 \pm
31$~\kms \citep{hmm+2012}, but we determined
$V_{HI}=6529 \pm 1$~\kms.


\subsection{Non-detected galaxies}
Limited by the observing time on the Parkes telescope, our observation
plan mainly focused on the galaxies which had a HIPASS peak flux
density larger than 20 mJy. Of the 303 observed galaxies, only 152
galaxies were well-detected with good spectra which meet the
requirements for accurate Tully-Fisher distance estimation. We
cross-matched the non-detected list with the HIPASS galaxy catalog,
and list these galaxies in Table~\ref{tab:non-det} for reference.
\begin{table}
\caption[]{Non-detected Galaxies}
\scriptsize
\label{tab:non-det}
\centering
\begin{tabular*}{0.98\columnwidth}{cccccc}
\hline
\hline
2MASX ID & RA (J2000) & DEC (J2000) & $V_{2MRS}$ & rms & Flag \\
 &[deg] &[deg] & [\kms] & [mJy]&\\
(1) & (2) & (3) & (4) & (5) & (6)\\
\hline
00011748-5300348	 &   0.3228	&  -53.0097	&9724	&5.20	&N\\
00032138-5004494 &   0.8390	&  -50.0805	&10333	&8.50	&N\\
00034062-4951278 &   0.9194	&  -49.8578	&8327	&7.12	&N\\
00054271-7542251 &   1.4278	&  -75.7070	&6028	&8.19	&N\\
00182593-8306394	 &   4.6081	&  -83.1110	&4534	&6.84	&Y\\
00254881-6219480 &   6.4533	&  -62.3300	&9174	&8.53	&N\\
00543231-4042578 &  13.6347	&  -40.7161	&7273	&7.84	&N\\
00571478-4057329	 &  14.3116	&  -40.9591	&3397	&7.40	&Y\\
01004798-5148563	 &  15.1999	&  -51.8156	&7449	&8.07	&N\\
01013572-5312020	 &  15.3988	&  -53.2005	&7457	&7.59	&N\\
01071459-4637191	 &  16.8109	&  -46.6220	&6081	&7.52	&Y\\
01093909-6119597	 &  17.4128	&  -61.3332	&7891	&3.79	&Y\\
01101993-4551184	 &  17.5830	&  -45.8551	&6968	&7.45	&Y\\
01281188-4334337	 &  22.0496	&  -43.5760	&9774	&7.94	&Y\\
01284236-5124573 &  22.1766	&  -51.4160	&9068	&7.10	&N\\
\hline

\end{tabular*}
\begin{tabular*}{0.98\columnwidth}{p{1.1cm}p{6.5cm}}
Notes --&\hspace{-0.48cm}Y: One or more peaks with flux $S_{peak} \geqslant
20$~mJy is found on the HIPASS spectrum in the velocity region of
$V_{2MRS}\pm200$~\kms.\\ &\hspace{-0.48cm}N: No peaks with flux $S_{peak}
\geqslant 20$~mJy are found on the HIPASS spectrum in the velocity
region of
$V_{2MRS}\pm200$~\kms.\\ \multicolumn{2}{p{0.95\columnwidth}}{Table~\ref{tab:non-det}
  is available in its entirety online.  A portion is shown here for
  guidance regarding its form and content.}\\
\end{tabular*}
\end{table}

\section{Summary}
\label{sec:conc}
We observed 303 galaxies in the southern hemisphere ($\delta <
-40^\circ$), as a part of the 2MASS Tully-Fisher survey, using the
Parkes radio telescope with the 21-cm multibeam receiver. The velocity
resolution of raw spectra is 1.6~\kms, after the 3 channel Hanning
smoothing during the data reduction process, The final velocity resolution
after Hanning smoothing is 3.3~\kms.  All galaxies were
selected from the 2MRS catalog with limits of
$K_{s}<11.25$~mag, $cz<10,000$~\kms, and axis ratio $b/a < 0.5$.

152 galaxies were detected with high quality spectra. We have
presented a table of both H\,{\sc i} spectral parameters and corrected
rotational velocities for these galaxies. All 152 galaxies have $\textrm{S/N} > 5$, and
66 have $\textrm{S/N} > 10$. We carefully measured the H\,{\sc i} spectral
parameters using a similar method to that applied to the 2MTF GBT and
Arecibo data, and converted the linewidths to rotational velocities,
which will be used for calculating the Tully-Fisher distances. 
We measured velocity widths with better than 5\% 
precision (suitable for application of the Tully-Fisher relation) 
for 148 out of 152 galaxies.

These observation comprise the southern portion of 2MTF and 
provide 69 high-accuracy measurements of galaxies in the southern Zone 
of Avoidance ($|b|<15^\circ$). The improved
uniformity and completeness will result in more accurate determinations of local peculiar
velocities.\\

We gratefully acknowledge help with Parkes Observations from 
John Huchra, Stacy Mader, A. Kels, Danny Price, Emma Kirby, Christina 
Magoulas and Vicky Safouris and all of the CSIRO staff at Parkes Observatory.
The authors wish particularly to acknowledge John Huchra (1948-2010), 
without whose vision 2MTF would never have happened. The 2MTF survey 
was initiated while KLM was a postdoc working with JPH at Harvard, and 
its design owes much to the insight and advice of JPH. 

Parts of this research were conducted by the Australian Research
Council Centre of Excellence for All-sky Astrophysics (CAASTRO),
through project number CE110001020.
TH was supported by the
National Natural Science Foundation (NNSF) of China
(10833003 and 11103032).
KLM was supported by NSF grant AST-0406906, the Peter and Patricia 
Gruber Foundation, and the Leverhulme Trust. 
LMM was supported by NASA through Hubble Fellowship grant 
HST-HF-01153 from the Space Telescope Science Institute and 
by the NSF through a Goldberg Fellowship from the National 
Optical Astronomy Observatory. ACC was supported by NSF grant AST-0406906.


\bibliographystyle{apj}
\bibliography{bibfile}

\begin{appendix}
\section{Errors in H\,{\sc i} parameters}
To estimate the errors on the H\,{\sc i} parameters, we used two different
methods.  A Monte-Carlo method was used for the errors of line widths
and central velocities, and a jackknife method was adopted for the
errors in flux. We describe these two methods in this section.  We also
compare the errors estimated by both methods with the errors estimated
by the method of the HIPASS Brightest Galaxy Catalog (HIPASS BGC
method).
\begin{figure}
\centering
\includegraphics[width=0.7\columnwidth, angle=-90]{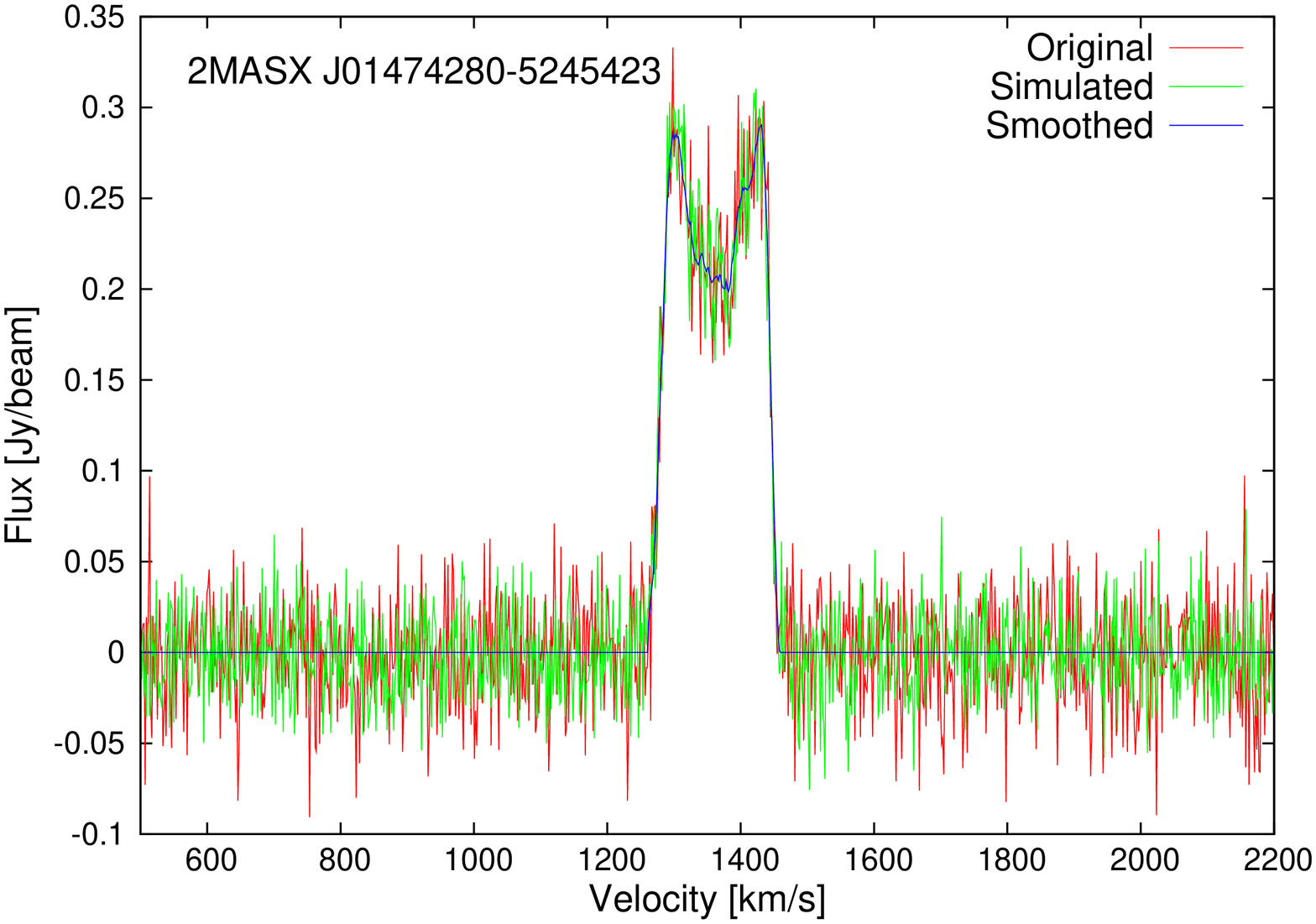}
\caption{The spectrum of galaxy 2MASX J01474280-5245423, together with
  the smoothed and mock spectrum. The red line indicates the original
  spectrum, the blue line shows the smoothed spectrum using a 17-point
  Savitzky-Golay smoothing filter, and the green line is the mock spectrum
  created by adding Poisson noise to the smoothed spectrum.}
\label{MC_error}
\end{figure}
\subsection{The Monte-Carlo method}
\label{sec:Monte}
We adopted a Monte-Carlo method to estimate the errors in the H\,{\sc i}
velocity parameters.  Firstly, we smoothed each galaxy spectrum using
a 17-point Savitzky-Golay smoothing filter \citep[\S 14.8]{ptvf2002}.
This low-pass filter can significantly reduce the noise while keeping
high order features of the spectrum. Fifty mock spectra were then
created for every galaxy by adding Poisson noise to the smoothed
spectrum. The rms of the random noise was equal to the rms of the
original galaxy spectrum. These mock spectra were measured with an
automatic IDL routine, based on the IDL routine \textit{awv.pro}. The
standard errors of the measurements of the fifty mock spectra were
then taken as the errors of the H\,{\sc i} parameters. Figure~\ref{MC_error}
shows the smoothed and mock spectrum of galaxy 2MASX J01474280-5245423
as an example.

\citet{dski+2005} adopted a similar analysis for the Parkes Zone of
Avoidance (ZoA) survey, and found this Monte-Carlo method worked well
for high S/N spectra while the errors became unreliable for
$\textrm{S/N} < 5$. For the 152 well-detected galaxies in our sample,
all galaxies have a peak $\textrm{S/N} > 5$. 86 have $5 < \textrm{S/N}
\leqslant 10$, and 66 have $\textrm{S/N} > 10$.

\begin{figure}
\centering
\includegraphics[width=0.8\columnwidth, angle=-90]{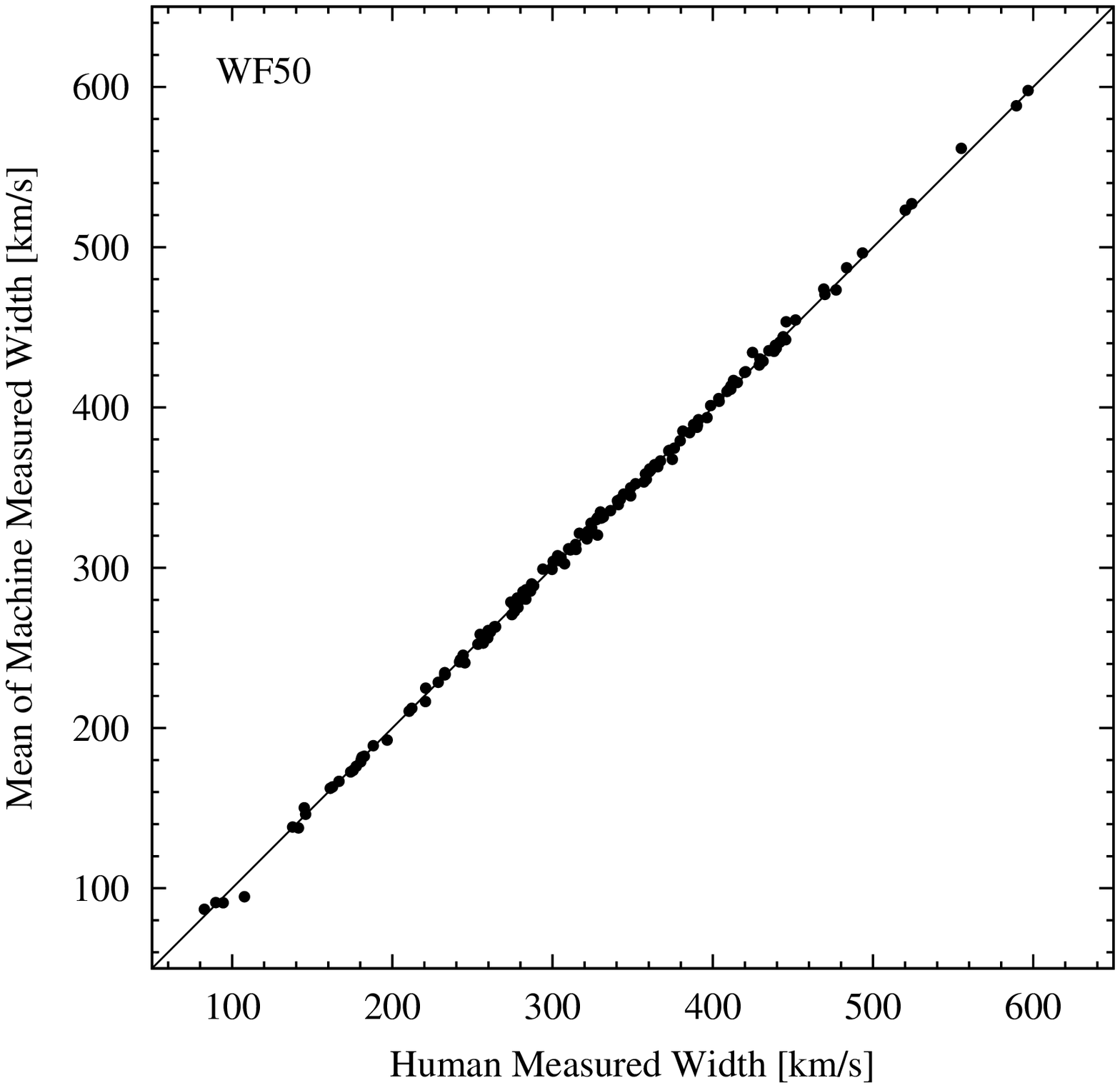}
\caption{The comparison plot of human-measured and
  machine-measured H\,{\sc i} $W_{F50}$ widths.  The solid line indicates
  equality. The scatter of these two measurements is about 2~\kms.}
\label{width_compare}
\end{figure}

\subsection{The jackknife method}
\label{sec:jack}
The Monte-Carlo method operates on the spectra following baseline
correction.  Since baseline correction is one of the major sources of error for the
measurement of H\,{\sc i} flux, we have adopted an alternative jackknife method to
estimate the errors in H\,{\sc i} flux.  After bandpass and Doppler correction
with \textsc{livedata}, we repeated the gridding and baseline fitting
process using an IDL routine instead of the \textsc{gridzilla}. As
mentioned in Section~\ref{sec:obs}, the correlator writes a spectrum
every 5 seconds for each polarization. Thus in a standard 35~mins
integration, 940 `sub-spectra' are recorded.

We built 100 jackknife spectra for every galaxy by removing 4
different polarization pairs of sub-spectra from the original data and
adding the rest of the spectra together using the MEDIAN method.

All the jackknife spectra were measured automatically using the same
IDL routines used in Section~\ref{sec:Monte}. Finally, we obtain the
jackknife estimate of the flux error from Equation~\ref{eq:sigma}.

\subsection{Reliability of the machine-measured H\,{\sc i} properties}
We compare the estimates of manual measurements with the mean value of
machine-measured H\,{\sc i} widths, to make sure our automatic routine can
measure the H\,{\sc i} profiles correctly. The comparison for our preferred
$W_{F50}$ widths is plotted in Figure~\ref{width_compare}, and shows
no significant systematic offset between manual and machine-measured
$W_{F50}$ widths.

\subsection{Comparison with the HIPASS BGC method}
\citet{ksk+2004} estimated the errors of the 1000 brightest HIPASS
galaxies using:

\begin{equation}
\label{eq:err_vel_HIP}
\sigma (v_{sys}) = 3 (S/N)^{-1} (P \Delta_v)^{1/2}
\end{equation} 
\begin{equation}
\label{eq:err_wid_HIP}
\sigma(w_{50})=2\sigma (v_{sys}),
\end{equation}
\begin{equation}
\label{eq:err_flux_HIP}
\sigma \left( F_{HI} \right) = 4 \left( S/N\right)^{-1} \left( S_{peak} F_{HI} \Delta_v \right)^{1/2},
\end{equation} 
where S/N is the signal-to-noise ratio, $S_{peak}$ is the peak flux
density, $\Delta_v=3.3$~\kms\ is the velocity resolution, and
$P=0.5(w_{20}-w_{50})$ indicates the slope of the H\,{\sc i} profile.

\begin{figure*}
\centering
\includegraphics[width=0.85\columnwidth, angle=-90]{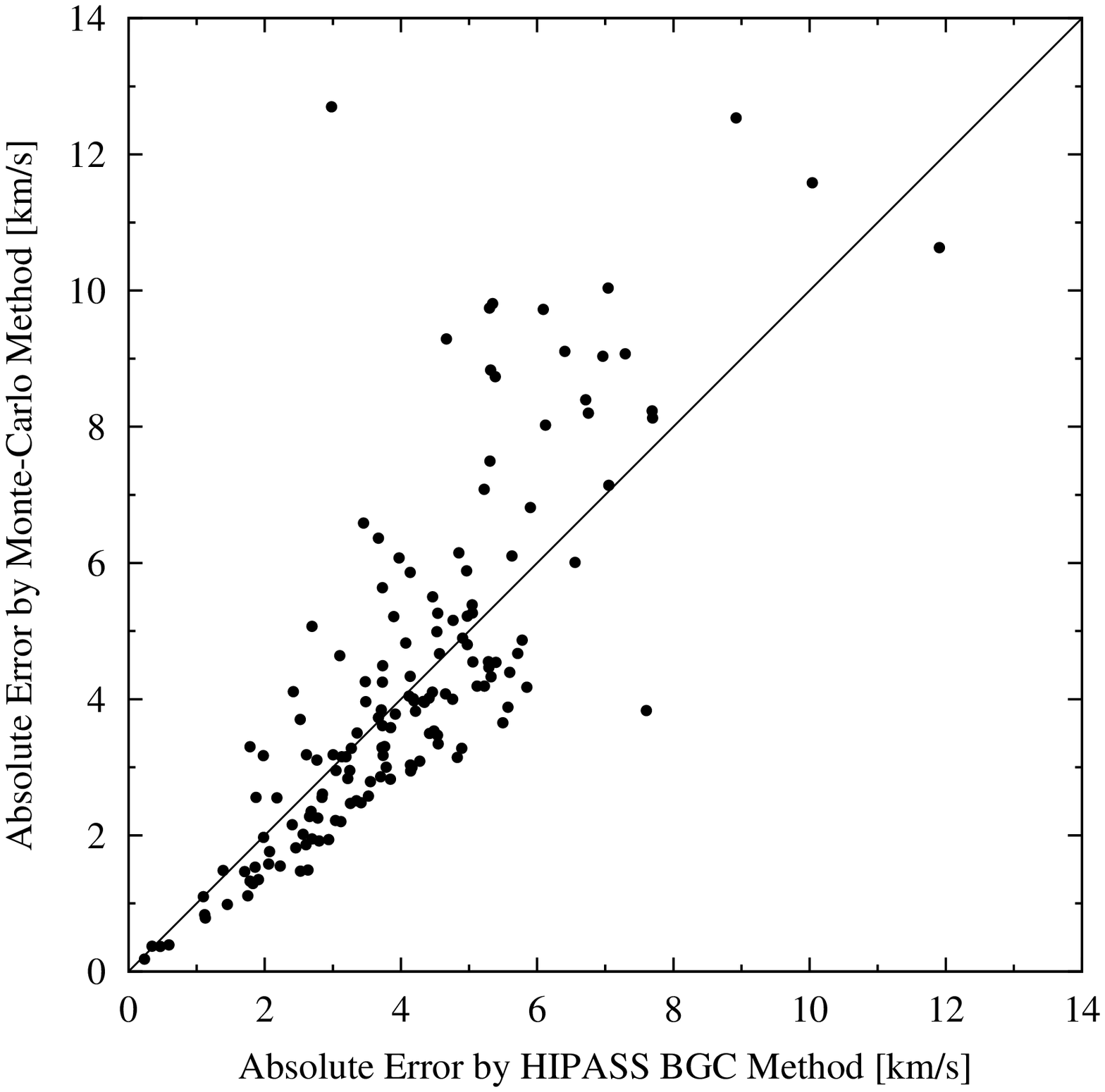}
\includegraphics[width=0.85\columnwidth, angle=-90]{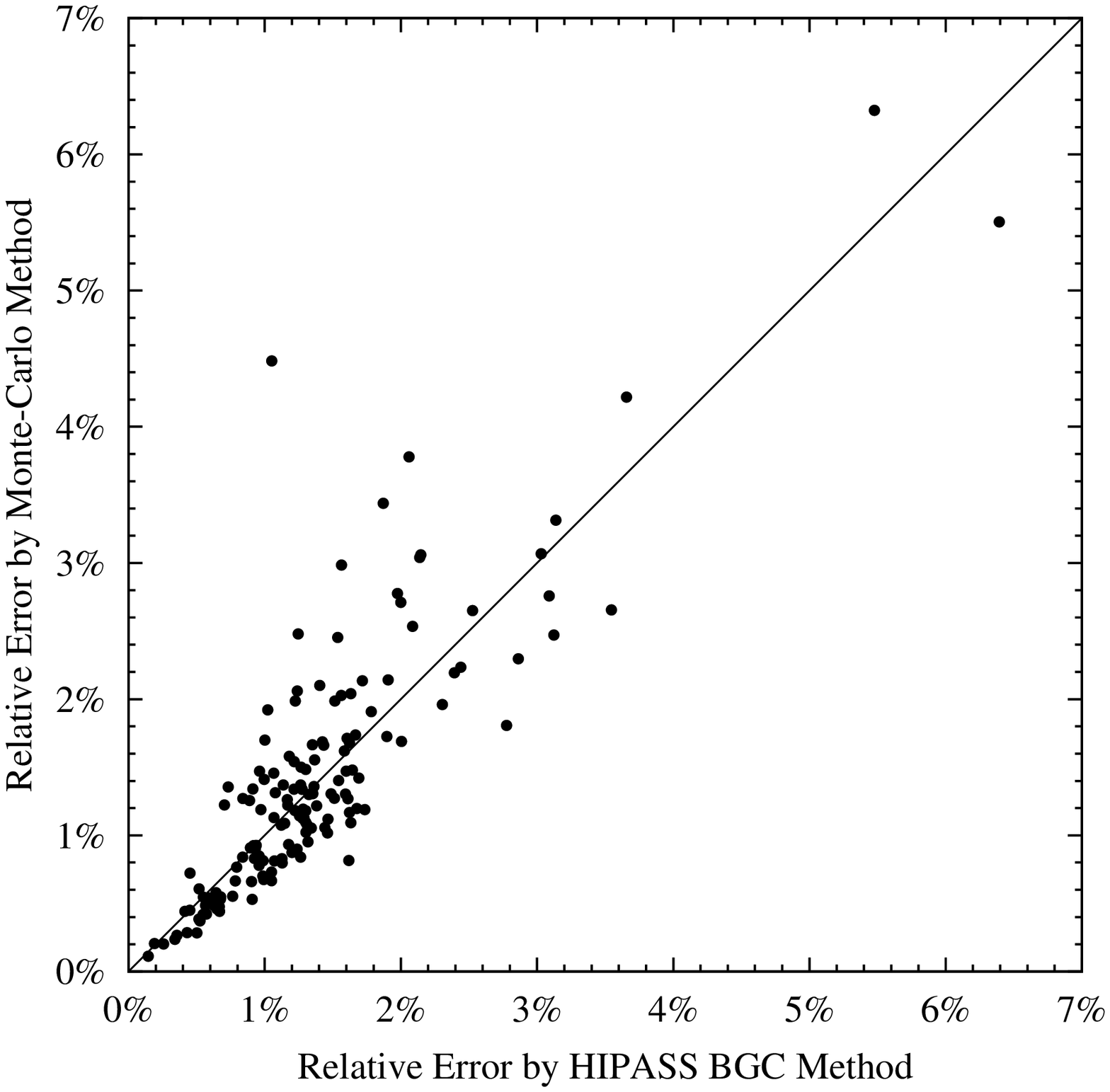}
\caption{Comparison of the Monte-Carlo width errors and the errors
  estimated using HIPASS BGC method. The solid line indicates the line
  of y=x.}
\label{fig:com_wid_HIP}
\end{figure*}

Firstly we compared the $W_{F50}$ width errors estimated by
the Monte-Carlo method with the errors calculated by
Equation~\ref{eq:err_wid_HIP} (Figure~\ref{fig:com_wid_HIP}). 
These two methods are consistent.  However, 
we find that the HIPASS BGC method tends to slightly overestimate the width
errors.
%

\begin{figure*}
\centering
\includegraphics[width=0.85\columnwidth, angle=-90]{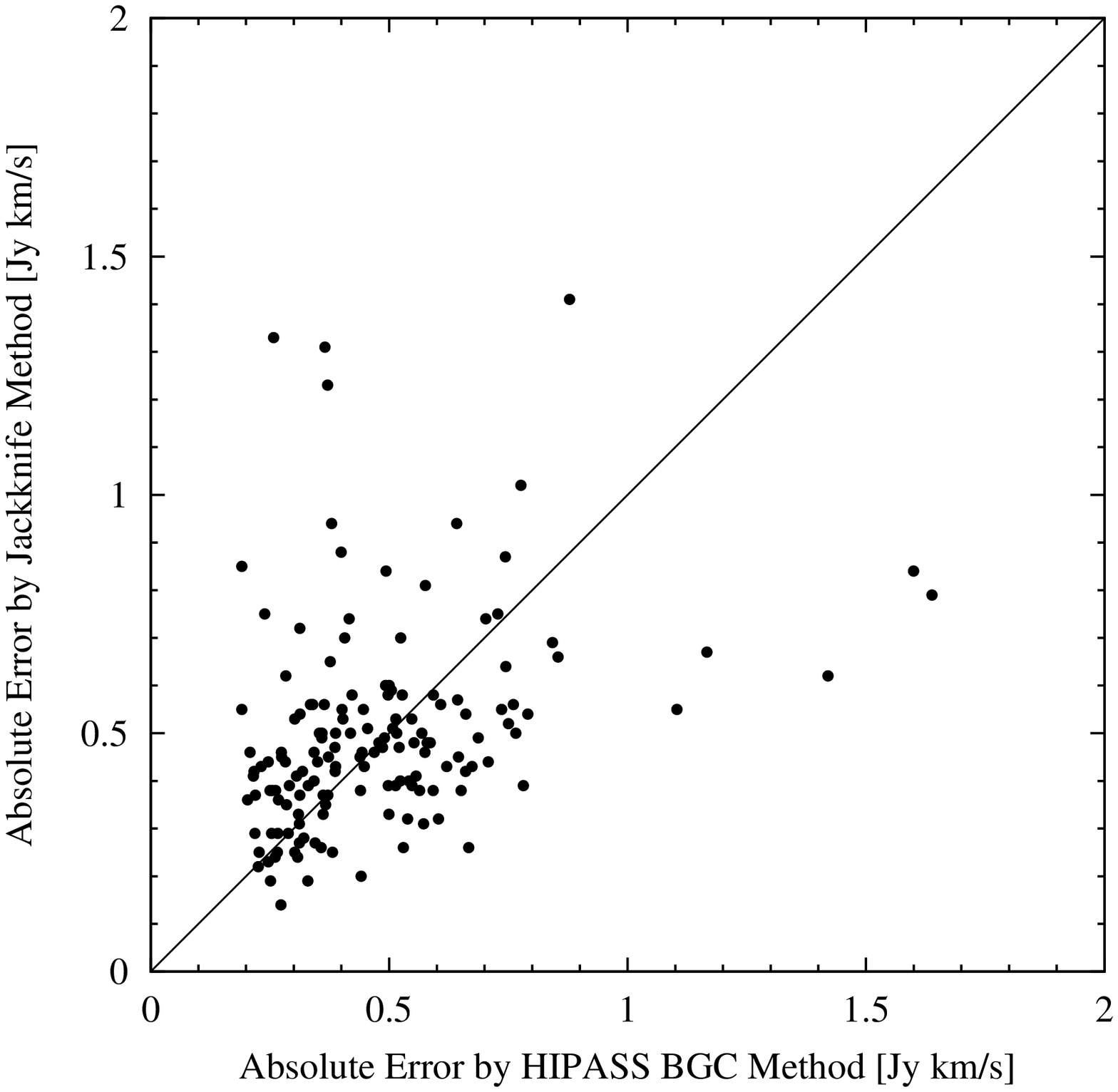}
\includegraphics[width=0.85\columnwidth, angle=-90]{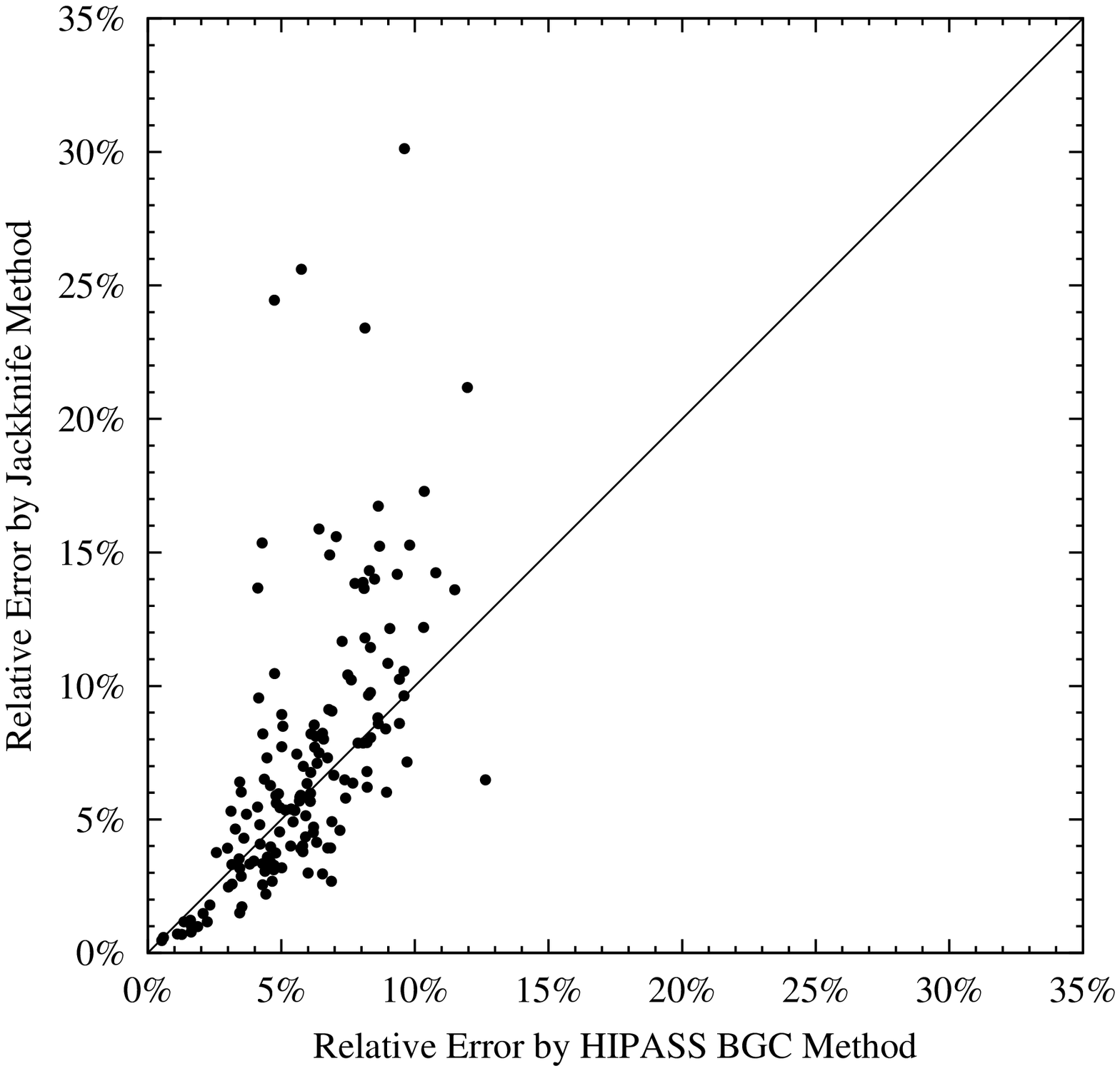}\\

\caption{Comparison of the jackknife flux errors and the errors
  estimated using the HIPASS BGC method. The solid line indicates the line
  of y=x.}
\label{fig:com_jac_HIP}
\end{figure*}

We also compared the flux error which was estimated using the
jackknife method with the errors of HIPASS BGC method (Figure~\ref{fig:com_jac_HIP}).
These two methods agree with each other, but with a large scatter,
especially for some low signal-to-noise ratio spectra. Our jackknife
method is more sensitive to the S/N than HIPASS BGC method, the
jackknife gave very large flux errors for low S/N galaxies. However,
for well observed galaxies, the two methods provide similar values.

\end{appendix}

\label{lastpage}
\end{document}